\def\lsim {~^{<~}_{\sim~}}
\def\gsim {~^{>~}_{\sim~}}

\documentclass{PoS}

\title{Analytical derivation of gauge fields from link variables in SU(3) lattice QCD and its application in maximally Abelian gauge}

\ShortTitle{Analytical derivation of gluons from link variables in SU(3)  and its application in MAG}

\author{\speaker{Shinya Gongyo}%
         \thanks{A footnote may follow.}\\
        Department of Physics, Kyoto University\\
        E-mail: \email{gongyo@ruby.scphys.kyoto-u.ac.jp}}
        
\author{Hideo Suganuma, Takumi Iritani\\
        Department of Physics, Kyoto University\\}

\abstract{In SU(3) lattice QCD, we improve a method to extract gauge fields 
from link variables analytically. With this method, we perform the first study 
on the effective mass generation of off-diagonal gluons and infrared Abelian 
dominance in the maximally Abelian (MA) gauge in the SU(3) case. 
We investigate the propagator and the effective mass of 
the off-diagonal gluon field in the MA gauge with 
${\rm U(1)}_3 \times {\rm U(1)}_8$ Landau gauge fixing 
in SU(3) quenched lattice QCD on $16^4$ at $\beta$=5.7, 5.8 and 6.0. 
The off-diagonal gluon component behaves as a massive vector boson 
with the approximate effective mass $M_{\rm off} = 1.1 - 1.2 {\rm GeV}$ 
in the region of $r = 0.3 - 0.8 {\rm fm}$, and its propagation is 
limited within a short range. 
We thus show the origin of infrared Abelian dominance in terms of 
short-range propagation of off-diagonal gluons.
We also investigate the functional form of the off-diagonal 
gluon propagator. We find that the functional form is well described by the 
four-dimensional Euclidean Yukawa-type function 
${\rm exp}\{-m_{\rm off} r\}/r$ with $m_{\rm off} = 1.3 -1.4 {\rm GeV}$ 
for $r = 0.1- 0.8{\rm fm}$. 
This also indicates that the spectral function has a negative region.
}

\FullConference{The 30 International Symposium on Lattice Field Theory - Lattice 2012,\\
		June 24-29, 2012\\
		Cairns, Australia}

\begin{document}
\section{Introduction}

For the quark-confinement mechanism, Nambu, 't Hooft and Mandelstam proposed the dual-superconductor picture \cite{N74}. This picture is based on the electromagnetic duality and the analogy with the one-dimensional squeezing of the magnetic flux in the type-II superconductor. In this picture, there occurs color-magnetic monopole condensation, and then the color-electric flux between the quark and the antiquark is squeezed as a one-dimensional tube due to the dual Higgs mechanism.  From the viewpoint of the dual-superconductor picture in QCD, however, there are two assumptions of Abelian dominance \cite{tH81,EI82} and monopole condensation.
Here, Abelian dominance means that only the diagonal gluon component plays the dominant role for the nonperturbative QCD phenomena like confinement.

The maximally abelian (MA) gauge has mainly been investigated from the viewpoint of the dual-superconductor picture \cite{KSW87,SY90,BWS91,SNW94,SST95,Mi95,Wo95,AS98,IH99,SAI02,BC03,Ko11} and the various lattice QCD Monte Carlo simulations show that the MA gauge fixing seems to support these assumptions \cite{KSW87,SY90,BWS91,SNW94,Mi95,Wo95,AS98,IH99,SAI02,BC03}.

According to these studies, the diagonal gluons seem to be significant to the infrared QCD physics, which is called ``infrared Abelian dominance". Infrared Abelian dominance means that off-diagonal gluons do not contribute to infrared QCD. Therefore, the essence of infrared Abelian dominance is the behavior of  the off-diagonal gluon propagator.

The gluon propagators in the MA gauge has been investigated in SU(2) lattice Monte Carlo simulations \cite{AS98, BC03,Cu01}.
 To investigate the gluon propagators in the MA gauge, it is desired to extract the gluons exactly from the link-variables, because the link-variable cannot be expanded even for a small lattice spacing due to large fluctuation of gluons. In SU(2) lattice case, the extraction is easy to be done without any approximation, because of the SU(2) property.
 With this extraction, the SU(2) lattice simulation suggests that the off-diagonal gluons do not propagate in the infrared region due to the effective mass $M_{\rm off} \simeq 1.2{\rm GeV}$, while the diagonal gluon widely propagates \cite{AS98}.
 
In this paper, we propose a method to extract the gluons from the link-variable directly and generally in SU(3) lattice QCD, and to investigate the gluon propagators in the MA gauge.

\section{SU(3) lattice  QCD results for gluon propagators in the MA gauge}
\label{3-1}
To begin with, we consider a useful and general method to extract the gauge fields analytically and exactly from the link-variables whether $|agA_\mu (x)| \ll 1$ is satisfied or not \cite{FN04,GIS12}.

In this method, each link-variable $U \equiv U_\mu(s)$ is diagonalized with a unitary matrix $\Omega$,

\begin{eqnarray}
	U _d &\equiv& \Omega U \Omega ^\dagger 
          = e^{i ag\Omega A \Omega ^\dagger } 
	\equiv
		\left( 
	\begin{array}{ccc}
		e^{i \theta _1} &  & 0 \\
		 & e^{i \theta _2} &  \\
		0 &  & e^{i \theta _3} \\
	\end{array} 
	\right) ,
\end{eqnarray}	
where $-\pi  \leq \theta _i < \pi $ $(i=1,2,3)$ is taken. Note that $\theta _1 + \theta _2 + \theta _3 = 0~({\rm mod}~ 2\pi)$ according to ${\rm Tr}\Omega A \Omega ^\dagger ={\rm Tr}A = 0~({\rm mod}~ 2\pi)$. This property is also numerically checked in the MA gauge with U(1)$_3 \times$U(1)$_8$ Landau gauge fixing.

We can derive gluon fields $A$ by taking the logarithm of $U_d$, 
\begin{eqnarray}
	\Omega A \Omega ^\dagger =
			\frac{1}{ag}\left( 
	\begin{array}{ccc}
		 \theta _1 &  & 0 \\
		 &  \theta _2 &  \\
		0 &  &  \theta _3 \\
	\end{array} 
	\right) \> \>
	\Rightarrow A = \frac{1}{ag}\Omega ^\dagger
	\left (
	\begin{array}{ccc}
		 \theta _1 &  & 0 \\
		 &  \theta _2 &  \\
		0 &  &  \theta _3 \\
	\end{array} 
	\right)
	\Omega. 
\end{eqnarray}
This formalism is quite general, because the derivation is correct in any gauge and even without any gauge fixing.

Using the SU(3) lattice QCD, we calculate the gluon propagators \cite{GIS12} in the MA gauge with the U(1)$_3\times$U(1)$_8$ Landau gauge fixing. In the MA gauge, to investigate the gluon propagators, we use the above-mentioned method. The Monte Carlo simulation is performed  with the standard plaquette action on the $16^4$ lattice with $\beta$ =5.7, 5.8 and 6.0 at the quenched level. All measurements are done every 500 sweeps after a thermalization of 10,000 sweeps using the pseudo heat-bath algorithm. We prepare 50 gauge configurations for the calculation at each $\beta$. The statistical error is estimated with the jackknife method.

Here, we study the Euclidean scalar combination of 
the diagonal (Abelian) and off-diagonal gluon propagators as
\begin{eqnarray}
G_{\mu\mu}^{\rm Abel}(r) &\equiv & \frac{1}{2} \sum_{a= 3,8} \left< A_\mu^a(x)A_\mu^a(y)\right>, \nonumber \\
				\label{eqn:AAf002}
G_{\mu\mu}^{\rm off}(r) &\equiv &
\frac{1}{6} \sum_{a\neq 3,8} \left< A_\mu^a(x)A_\mu^a(y)\right>.
				\label{eqn:AAf003}
\end{eqnarray}
The scalar combination of the propagator is expressed as the function of the four-dimensional Euclidean distance $r\equiv \sqrt{(x_\mu -y_\mu)^2}$. 
When we consider the renormalization, 
these propagators are multiplied by an $r$-independent constant, according 
to a constant renormalization factor of the renormalized gluon fields.

\begin{figure}[h]
\begin{center}
\includegraphics[scale=0.6]{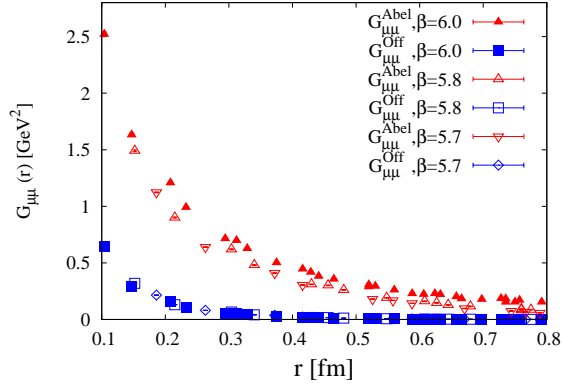}
\caption{
The SU(3) lattice QCD results of the gluon propagators $G_{\mu\mu}^{\rm Abel}(r)$ and $G_{\mu\mu}^{\rm off}(r)$ as the function of $r\equiv \sqrt{(x_\mu -y_\mu)^2}$ in the MA gauge with the U(1)$_3\times$U(1)$_8$ Landau gauge fixing in the physical unit. The Monte Carlo simulation is performed on the $16^4$ lattice with $\beta$ = 5.7, 5.8 and 6.0. The diagonal-gluon propagator $G_{\mu\mu}^{\rm Abel}(r)$ takes a large value even at the long distance. On the other hand, the off-diagonal gluon propagator $G_{\mu\mu}^{\rm off}(r)$ rapidly decreases.
}
\label{Fig1}
\end{center}
\end{figure}

We show in Fig.\ref{Fig1} the lattice QCD result for the diagonal gluon propagator $G_{\mu\mu}^{\rm Abel}(r)$ and the off-diagonal gluon propagator $G_{\mu\mu}^{\rm off}(r)$ in the MA gauge with the U(1)$_3\times$U(1)$_8$ Landau gauge fixing. In the MA gauge, $G_{\mu\mu}^{\rm Abel}(r)$ and 
$G_{\mu\mu}^{\rm off}(r)$ manifestly differ.
The diagonal-gluon propagator $G_{\mu\mu}^{\rm Abel}(r)$ 
takes a large value even at the long distance. 
In fact, the diagonal gluons $A_\mu^3,A_\mu^8$ in the MA gauge 
propagate over the long distance.
In contrast, the off-diagonal gluon propagator 
$G_{\mu\mu}^{\rm off}(r)$ rapidly decreases and is negligible
for $r \gsim 0.4$fm in comparison with $G_{\mu\mu}^{\rm Abel}(r)$. 
Then, the off-diagonal gluons $A_\mu^a~(a\neq 3,8)$ seem to propagate 
only within the short range as $r \lsim 0.4$fm.
Thus, ``infrared abelian dominance" is found in the MA gauge.

\section{Estimation of off-diagonal gluon mass in the MA gauge}
\label{3-2}
Next, we investigate the effective mass of off-diagonal gluons \cite{GIS12}.
We start from the Lagrangian of 
the free massive vector field $A_\mu$ with the mass $M \ne 0$ 
in the Euclidean metric. In the infrared region with large $Mr$, 
the propagator $G_{\mu\mu}(r;M)$ reduces to 
\begin{eqnarray}
{G}_{\mu\mu}(r;M) &=& \left< A_\mu(x) A_\mu(y) \right> 
	\simeq
	\frac{3\sqrt{M}}{2(2\pi)^{\frac{3}{2}}} \frac{e^{-Mr}}{r^\frac{3}{2}}, 
						\label{eqn:prp03} 
\end{eqnarray}
\begin{figure}[h]
\begin{center}
\includegraphics[scale=0.4]{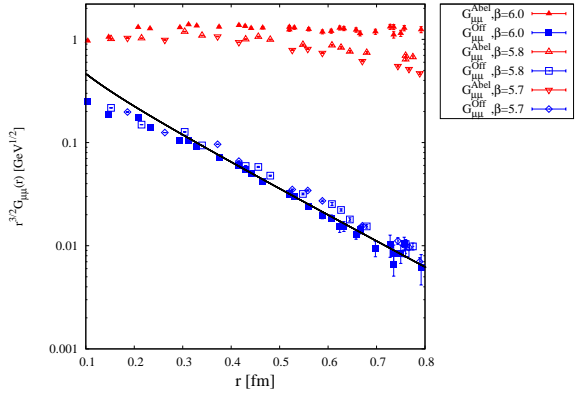}
\caption{
The logarithmic plot of $r^{3/2}G_{\mu\mu}^{\rm off} (r)$ and $r^{3/2}G_{\mu\mu}^{{\rm Abel}} (r)$ as the function of the Euclidean distance $r$ in the MA gauge with the U(1)$_3\times$U(1)$_8$ Landau gauge fixing, in the SU(3) lattice QCD with $16^4$ at $\beta$ = 5.7, 5.8 and 6.0. The solid line denotes the logarithmic plot of $r^{3/2}G_{\mu\mu}(r) \sim r^{1/2}K_1 (Mr)$ in the Proca formalism.
}
\label{Fig2}
\end{center}
\end{figure}

In Fig.\ref{Fig2}, we show the logarithmic plot of $r^{3/2}G_{\mu\mu}^{\rm off} (r)$ and $r^{3/2}G_{\mu\mu}^{\rm Abel} (r)$ as the function of the Euclidean distance $r$ in the MA gauge with the U(1)$_3\times$U(1)$_8$ Landau gauge fixing. From the linear slope on $r^{3/2}G_{\mu\mu}^{\rm off} (r)$ in the range of $r=0.3-0.8~{\rm fm}$, the effective off-diagonal gluon mass $M_{\rm off}$ is estimated. 
Note that the gluon-field renormalization does not affect the gluon mass 
estimate, since it gives only an overall constant factor for the propagator. 
We summarize in Table~1 the effective off-diagonal gluon mass $M_{\rm off}$ 
obtained from the slope analysis 
at $\beta$ =5.7, 5.8 and 6.0.
Therefore, the off-diagonal gluons seem to have a large mass
$M_{\rm off} \simeq 1.1-1.2~{\rm GeV}$.
This result approximately coincides with SU(2) lattice calculation \cite{AS98}.

\begin{table}[h]
\caption{Summary table of conditions and results in SU(3) lattice QCD. In the MA gauge, the off-diagonal gluons seem to have a large effective mass $M_{\rm off} \simeq 1.1-1.2~\mathrm{GeV}$ and the functional form in the range of $r=0.1-0.8~{\rm fm}$ is well described with the four-dimensional Euclidean Yukawa function $\sim \exp(-m_{\rm off}r)/r$  with $m_{\rm off} \simeq 1.3-1.4~\mathrm{GeV}$. }
\begin{center} 
\begin{tabular}{ccccc}
\hline \hline
  lattice size   & $\beta$     & $a[{\rm fm}]$ &   $M_{\rm off} [{\rm GeV}] $  & $m_{\rm off} [{\rm GeV}] $ \\
\hline
                 &    5.7  & 0.186 &  1.2  & 1.3 \\
$16^4$                 &    5.8  & 0.152 &  1.1 & 1.3 \\
                 &    6.0     &  0.104 & 1.1 & 1.4 \\
\hline \hline
\end{tabular}
\end{center} 
\end{table}
Finally in this section, we discuss the relation 
between infrared abelian dominance and 
the off-diagonal gluon mass.
Due to the large effective mass $M_{\rm off} $, 
the off-diagonal gluon propagation is restricted within about
$M_{\rm off}^{-1} \simeq 0.2$fm in the MA gauge.
Therefore, at the infrared scale as $r \gg 0.2{\rm fm}$,
the off-diagonal gluons $A_\mu^a~(a\neq 3,8)$ cannot mediate the long-range force like the massive weak bosons in the Weinberg-Salam model, 
and only the diagonal gluons $A_\mu^3,~A_\mu^8$ can mediate  
the long-range interaction in the MA gauge.
In fact, in the MA gauge, the off-diagonal gluons are expected to be 
inactive due to the large mass $M_{\rm off}$ in the infrared region 
in comparison with the diagonal gluons. 
Then, infrared abelian dominance holds for $r \gg M^{-1}_{\rm off}$. 

\section{Analysis of the functional form of off-diagonal gluon propagator in the MA gauge}
\label{3-3}
In this section, we investigate the functional form of the off-diagonal gluon propagator $G_{\mu\mu}^{\rm off}(r)$ in the MA gauge in SU(3) lattice QCD \cite{GIS12}. In the previous section, we compare the off-diagonal gluon propagator with the massive vector boson propagator and estimate the gluon mass. In fact, the gluon propagator would not be described by a simple massive propagator in the whole region of $r=0.1-0.8~{\rm fm}$.

There is the similar situation in the Landau gauge \cite{IS09}. The functional form of the gluon propagator cannot be described by $\exp(-Mr)/r^{3/2}$ with an effective mass $M$ in the whole region of $r=0.1-1.0~{\rm fm}$. The appropriate form is the four-dimensional Euclidean Yukawa-type function $\exp(-mr)/r$ with a mass parameter $m$.

\begin{figure}[h]
\begin{center}
\includegraphics[scale=0.5]{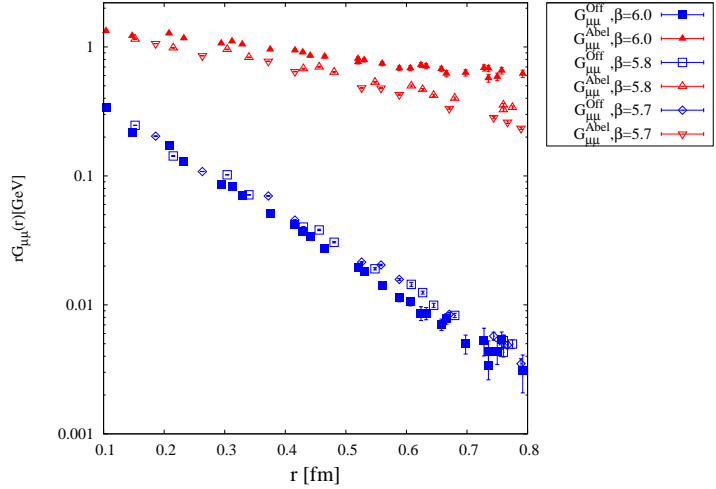}
\caption{
The logarithmic plot of $rG_{\mu\mu}^{\rm off} (r)$ and $rG_{\mu\mu}^{{\rm Abel}} (r)$ as the function of the Euclidean distance $r$ in the MA gauge with the U(1)$_3\times$U(1)$_8$ Landau gauge fixing, using the SU(3) lattice QCD with $16^4$ at $\beta$=5.7, 5.8 and 6.0. For $rG_{\mu\mu}^{\rm off} (r)$, the approximate linear correlaton is found.}
\label{Fig3}
\end{center}
\end{figure}

In the same way, in the MA gauge, we also compare the gluon propagator with the four-dimensional Euclidean Yukawa function.
 In Fig.\ref{Fig3}, we show the logarithmic plot of $rG_{\mu\mu}^{\rm off} (r)$ and $rG_{\mu\mu}^{\rm Abel} (r)$ as the function of the distance $r$ in the MA gauge with the U(1)$_3\times$U(1)$_8$ Landau gauge fixing.
Note that the logarithmic plot of $rG_{\mu\mu}^{\rm off} (r)$ is almost linear in the whole region of $r=0.1-0.8~{\rm fm}$, 
and therefore the off-diagonal gluon propagator is well expressed by the four-dimensional Euclidean Yukawa function $Ae^{-m_{\rm off} r}/r$, 
with a mass parameter $m_{\rm off}$ and a dimensionless constant $A$.
The best-fit mass parameter $m_{\rm off}$ is given in Table 1 at each $\beta$ = 5.7, 5.8 and 6.0. 

We comment on the four-dimensional Euclidean Yukawa-type propagator \cite{IS09}. If the functional form of the off-diagonal gluon is well described by the four-dimensional Yukawa function, we analytically calculate the off-diagonal zero-spatial-momentum propagator,
 $D_0^{\rm off}  (t)\equiv \int d^3x G_{\mu\mu}^{\rm off} (r)$, and obtain 
the spectral function by the inverse Laplace transformation. 
Similarly in the Landau gauge \cite{IS09}, we thus derive 
the spectral function $\rho^{\rm off}(\omega)$ 
of off-diagonal gluons in the MA gauge, 
\begin{eqnarray}
\rho^{\rm off} (\omega) = - \frac{4\pi Am_{\rm off}}{(\omega ^2 -m_{\rm off}^2)^{3/2}}\theta (\omega -m_{\rm off}) + \frac{4\pi A/\sqrt{2m_{\rm off}}}{(\omega - m_{\rm off})^{1/2}}\delta(\omega -m_{\rm off}).
\end{eqnarray} 

\section{Summary and Concluding Remarks}
\label{4}
We have performed the first study of the gluon propagators in the MA gauge 
with the U(1)$_3\times$U(1)$_8$ Landau gauge fixing in the SU(3) quenched 
lattice QCD. 
 To investigate the gluon propagators in the MA gauge, we have considered to derive the gluon fields analytically from the SU(3) link-variables.
 
With this method, we have calculated the Euclidean scalar combination 
$G_{\mu\mu} (r)$ of the diagonal and the off-diagonal gluon propagators, 
and have considered the origin of infrared Abelian dominance. 
The Monte Carlo simulation is performed on the $16^4$ lattice at 
$\beta$=5.7, 5.8 and 6.0 at the quenched level.
We have found that 
the off-diagonal gluons behave as massive vector bosons 
with the effective mass $M_{\rm off} \simeq 1.1-1.2$~GeV for $r =0.3-0.8$~fm. 
The effective gluon mass has been estimated from the linear fit analysis of the logarithmic plot of $r^{3/2}G_{\mu\mu} ^{\rm off}(r)$. 
Due to the large value, the finite-size effect for the off-diagonal gluon mass 
is expected to be ignored. The large gluon mass shows that the off-diagonal 
gluons cannot mediate the interaction over the large distance as 
$r \gg M_{\rm off}^{-1}$, and such an infrared inactivity of the off-diagonal 
gluons would lead infrared Abelian dominance in the MA gauge.

On the other hand, from the behavior of the diagonal gluon propagator 
$G_{\mu\mu} ^{\rm Abel}(r)$ and $r^{3/2}G_{\mu\mu} ^{\rm Abel}(r)$, 
 the diagonal gluons seem to behave as light vector bosons \cite{GIS12}.
For the detailed argument on $G_{\mu\mu}^{\rm Abel}(r)$,
one should consider the finite size effect more carefully,
because the diagonal gluons would propagate over the long distance 
beyond the lattice size.

Finally, we have also investigated the functional form of the off-diagonal 
gluon propagator $G_{\mu\mu} ^{\rm off}(r)$ in the MA gauge. 
We show that $G_{\mu\mu} ^{\rm off}(r)$ is well described by the 
four-dimensional Euclidean Yukawa-type form with the mass parameter 
$m_{\rm off} \simeq 1.3 -1.4$~GeV in the whole region of $r=0.1-0.8$~fm.
This indicates that the spectral function $\rho^{\rm off} (\omega)$ of 
the off-diagonal gluons in the MA gauge 
has the negative-value region \cite{GIS12}, 
as in the Landau gauge \cite{IS09,MO87,Bo04}.


In this study, we investigate the off-diagonal gluon propagator. To be strict, the off-diagonal gluon propagator consists of two scalar functions corresponding to longitudinal and transverse components. Therefore, we will investigate each effective mass and the functional form of these components.

\section*{Acknowledgements}
The authors are grateful to Dr. Hideaki Iida for useful discussions.
This work is supported in part by the Grant for Scientific Research 
[(C) No.~23540306, Priority Areas ``New Hadrons'' (E01:21105006)], Grant-in-Aid for JSPS Fellows (No.23-752, 24-1458)
from the Ministry of Education, Culture, Science and Technology 
(MEXT) of Japan, and the Global COE Program, 
``The Next Generation of Physics, Spun from Universality and Emergence".
The lattice QCD calculations are done on NEC SX-8R at Osaka University.

\end{document}